\documentclass[review]{elsarticle}
\usepackage[utf8]{inputenc}
\usepackage{fullpage}
\usepackage{amsmath}
\usepackage{mathtools}
\usepackage{setspace}
\usepackage{booktabs}
\usepackage[ruled]{algorithm2e}[1]
\usepackage{graphicx}
\usepackage{longtable}
\usepackage{amssymb}
\usepackage{float}
\usepackage{soul}
\usepackage[dvipsnames]{xcolor}
\usepackage{algpseudocode}  

\journal{Journal of \LaTeX\ Templates}

\begin{document}

\begin{frontmatter}
\title{Partial domain adaptation enables cross domain cell type annotation between scRNA-seq and snRNA-seq}


\cortext[cor1]{Corresponding author}

\author[Shandong,sichuan,jiaotong]{Xiran Chen} 
\author[sichuan]{Quan Zou}
\author[westlake]{Qinyu Cai}
\author[jiaotong]{Xiaofeng Chen}
\author[Shandong]{Weikai Li \corref{cor1} }
\author[sichuan]{Yansu Wang \corref{cor1} }

\affiliation[Shandong]{organization = {School of Computer and Artificial Intelligence, Shandong Jianzhu University, Shandong, China}}
\affiliation[sichuan]{organization={Institute of Fundamental and Frontier Sciences, University of Electronic Science and Technology of China, Chengdu, Sichuan, China}}
\affiliation[jiaotong]{organization={School of Mathematics and Statistics, Chongqing Jiaotong University, Chongqing, China}}
\affiliation[westlake]{organization = {School of Life Sciences, Westlake University, Hangzhou, China}}

\begin{abstract}

Accurate cell type annotation across datasets is a key challenge in single-cell analysis. snRNA-seq enables profiling of frozen or difficult-to-dissociate tissues, complementing scRNA-seq by capturing fragile or rare cell types. However, cross-annotation between these two datasets remains largely unexplored, as existing methods treat them independently. We introduce ScNucAdapt, a method designed for cross-annotation between paired and unpaired scRNA-seq and snRNA-seq datasets. To address distributional and cell composition differences, ScNucAdapt employs partial domain adaptation. Experiments across both unpaired and paired scRNA-seq and snRNA-seq show that ScNucAdapt achieves robust and accurate cell type annotation, outperforming existing approaches. Therefore, ScNucAdapt provides a practical framework for the cross-domain cell type annotation between scRNA-seq and snRNA-seq data.
\end{abstract}


\end{frontmatter}
\section*{Author Summary}
Single-cell and single-nucleus RNA sequencing are two powerful technologies that allow scientists to study gene activity in individual cells. However, comparing data between these methods remains challenging because they capture different parts of the cell and are often collected under different conditions. This makes it difficult to consistently identify cell types across experiments, hindering our understanding of health and disease.

We developed ScNucAdapt, a computational framework that can automatically transfer cell type knowledge between these two types of datasets, even when they come from different laboratories or tissue conditions. Our method learns to recognize shared patterns while ignoring dataset-specific differences. Through testing on diverse tissues, including bladder, kidney, tumors, and brain, we show that ScNucAdapt consistently outperforms existing approaches.

By enabling reliable integration of single-cell and single-nucleus data, our work helps researchers build more complete pictures of cellular diversity across tissues and disease states. This capability is particularly valuable for studying archived frozen samples or fragile cell types that are difficult to analyze with conventional methods, potentially accelerating discoveries in various fields.
\section{Introduction}
Cells are the fundamentals of life  \cite {regev2017human}. It is vital to annotate each cell correctly in terms of its transcriptomic profiles  \cite {pasquini2021automated}, enabling the identification of distinct cellular populations, comparison across samples, and linkage of molecular profiles to biological function or disease \cite {li2025overview}. Most published methods on automatic cell type annotations are based on scRNA-seq, including SingleCellNet \cite {tan2019SingleCellNet}, which uses an ensemble of Random Forest classifiers for the cell type annotation of scRNA-seq datasets, and ScMap \cite {kiselev2018scmap} works by comparing the gene-expression profile of each new cell to reference data and labeling the cell with the type that shows the highest similarity. 

However, when confronted with frozen samples or tissues that are difficult to dissociate, snRNA-seq offers a practical alternative to scRNA-seq \cite {ding2020systematic}, capturing nuclear transcripts without viable whole cells and enabling detection of fragile or rare cell types that single-cell methods often underrepresent \cite {wu2019advantages}. Many studies have integrated scRNA-seq and snRNA-seq, for instance, neurodegenerative diseases \cite {zhang2023comprehensive}, skeletal muscle \cite {heuston2025optimized}, frozen and fresh tumor samples \cite {slyper2020single}, and even PBMC for a disease progression study \cite {park2025paired}. This shows that cross-domain annotation between scRNA-seq and snRNA-seq is essential for unifying cellular identities across two datasets and ensuring consistent interpretation of data generated from different tissue conditions or experimental protocols \cite {quatredeniers2023meta}. Previous research on annotating cell types for snRNA-seq and scRNA-seq data relies on traditional machine learning methods \cite {water2}, particularly for kidney cell types \cite {water}. However, these methods overlooked the relationships between scRNA-seq and snRNA-seq, treating them as separate datasets. Therefore, there's an urgent need for development in cross-domain cell type annotation between scRNA-seq and snRNA-seq.

\begin{figure}[H]
\includegraphics[width=1\textwidth]{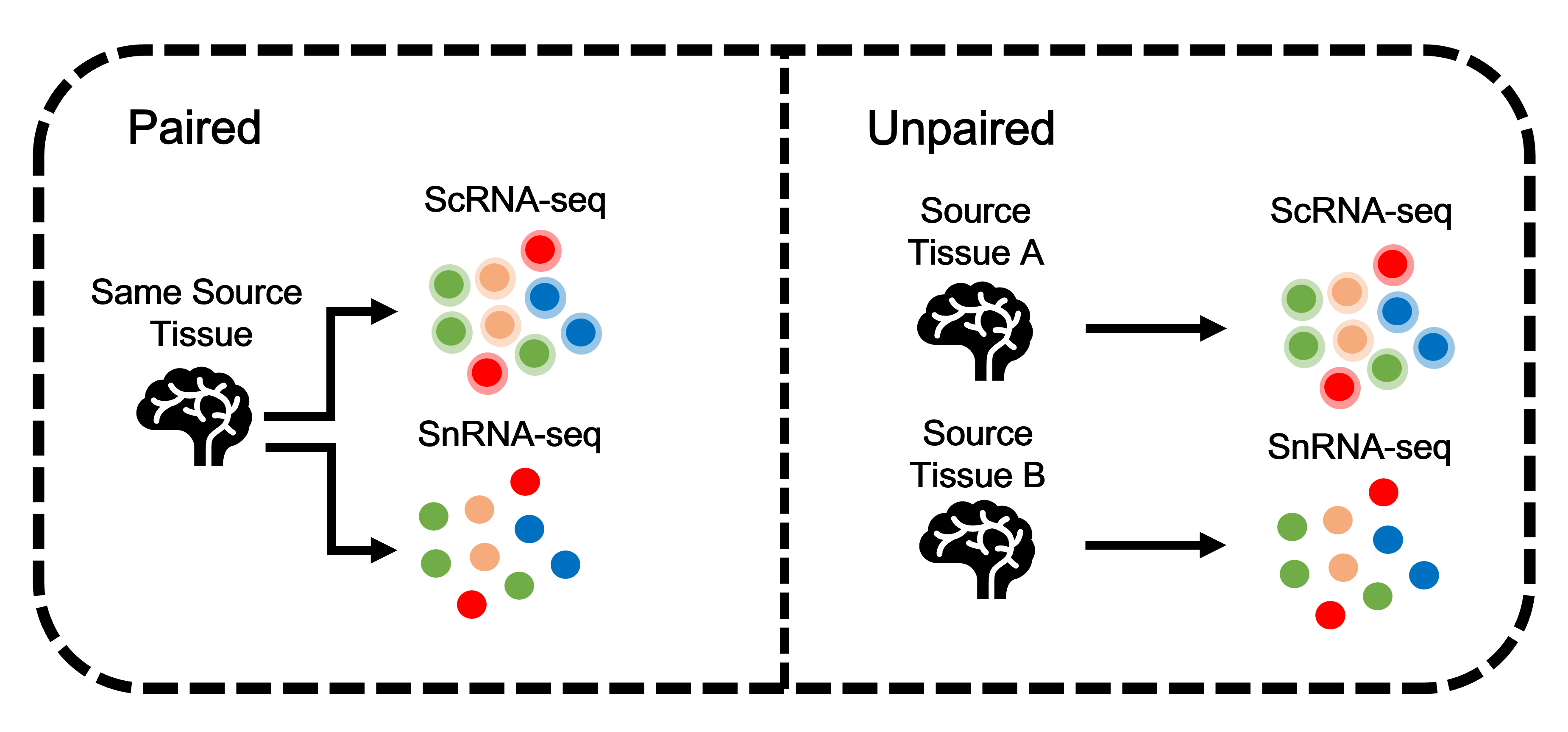}
\caption{Concept of pair and unpaired scRNA-seq and snRNA-seq}\label{pair_unpair}
\end{figure}
However, Distributional differences often occur between scRNA-seq and snRNA-seq \cite {andrews2022single}, including paired and unpaired ones, as shown in Figure \ref{pair_unpair}. Moreover, in a real automatic annotation situation, the cell type composition of target datasets is unknown, which could cause cell type compositions to differ between the two datasets, making it challenging to achieve robust annotation between them. Therefore, inspired by partial domain adaptation, which can simultaneously address both the distribution differences between the two datasets and the mismatch in their label spaces, we developed a framework called ScNucAdapt that selectively transfers knowledge from the source dataset to the target dataset.

\begin{figure}[H]
\includegraphics[width=1\textwidth]{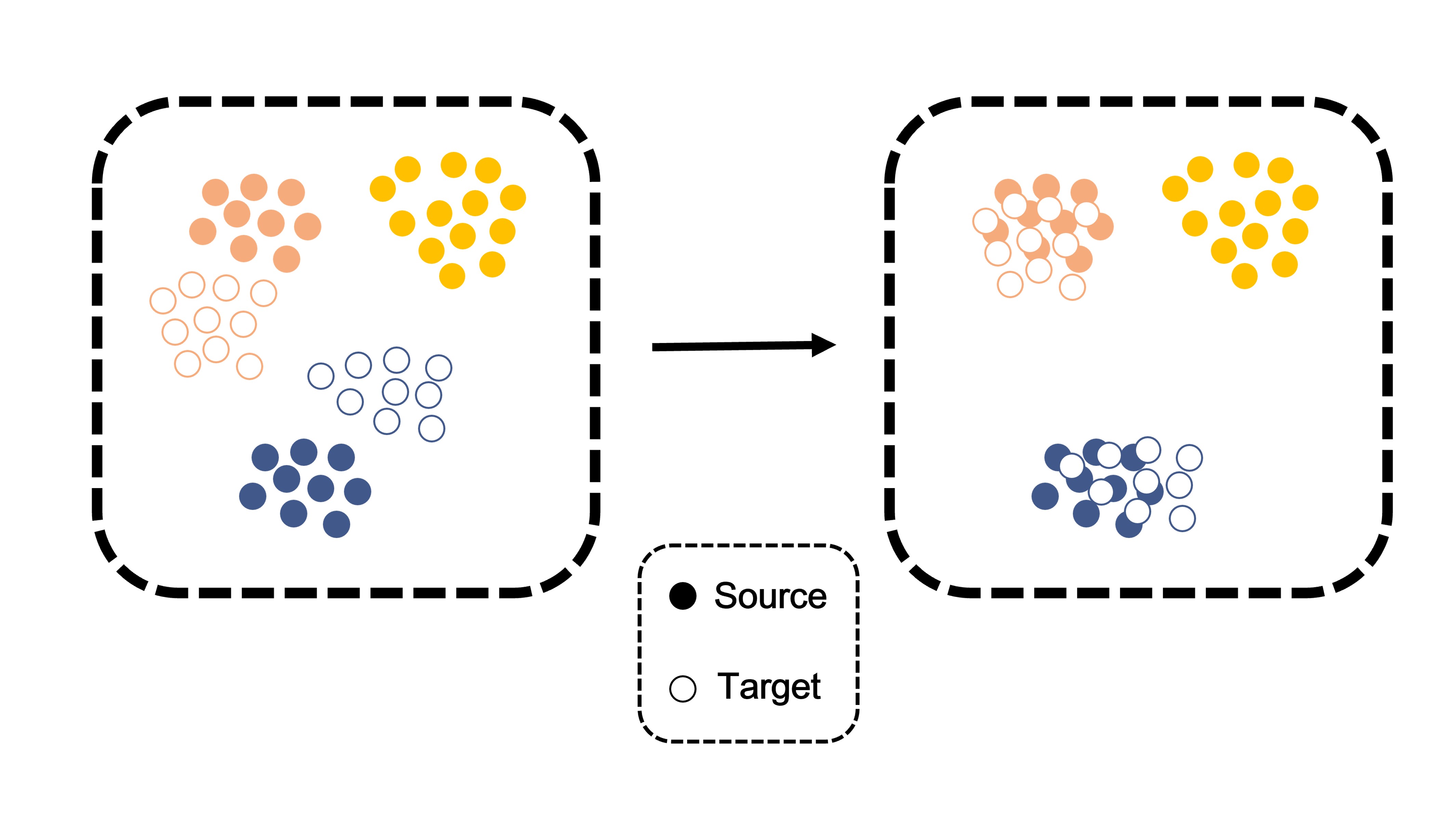}
\caption{Concept of partial domain adaptation}\label{partial}
\end{figure}

Partial domain adaptation \cite {li2022partial} addresses the problem of transferring knowledge from a labeled source domain to an unlabeled target domain when the target label space is a subset of the source label space, as shown in Figure \ref{partial}. Unlike traditional domain adaptation, which assumes identical label spaces across domains, Partial domain adaptation mitigates the negative transfer caused by irrelevant source classes. Partial domain adaptation has been applied to various fields, including fault diagnosis \cite {zhu2023partial}, pneumonia diagnosis from chest x-ray images \cite {liu2023attention}, and cross-session neural decoding \cite {wang2025partial}.

This design enables ScNucAdapt to focus on cell types that are shared across datasets while minimizing the negative impact of non-overlapping or dataset-specific cell types. Moreover, ScJoint \cite {scjoint}, ScNCL \cite {yan2023scncl}, ScCobra \cite {zhao2025sccobra}, and ScCorrect \cite {liu2025sccorrect}, which utilize transfer learning and domain adaptation methods for label transfer from unpaired scRNA-seq to ScATAC-seq datasets, also provided inspiration for developing ScNucAdapt.

To the best of our knowledge, Our study is the first to focus on cross-annotation between paired or unpaired scRNA-seq and snRNA-seq datasets. In summary, the contributions of our proposed method can be listed as follows:
\begin{itemize}
    \item ScNucAdapt is a cross-domain annotation framework that enables robust label transfer between paired or unpaired scRNA-seq and snRNA-seq datasets.
    \item ScNucAdapt also considers distributional differences between scRNA-seq and snRNA-seq, making it more robust when annotating cell types in the target datasets.
    \item ScNucAdapt also could handle cell type compositional differences between scRNA-seq and snRNA-seq, where only a subset of cell types are shared across these two datasets.
\end{itemize}

\begin{figure}[H]
    \includegraphics[width=1\textwidth]{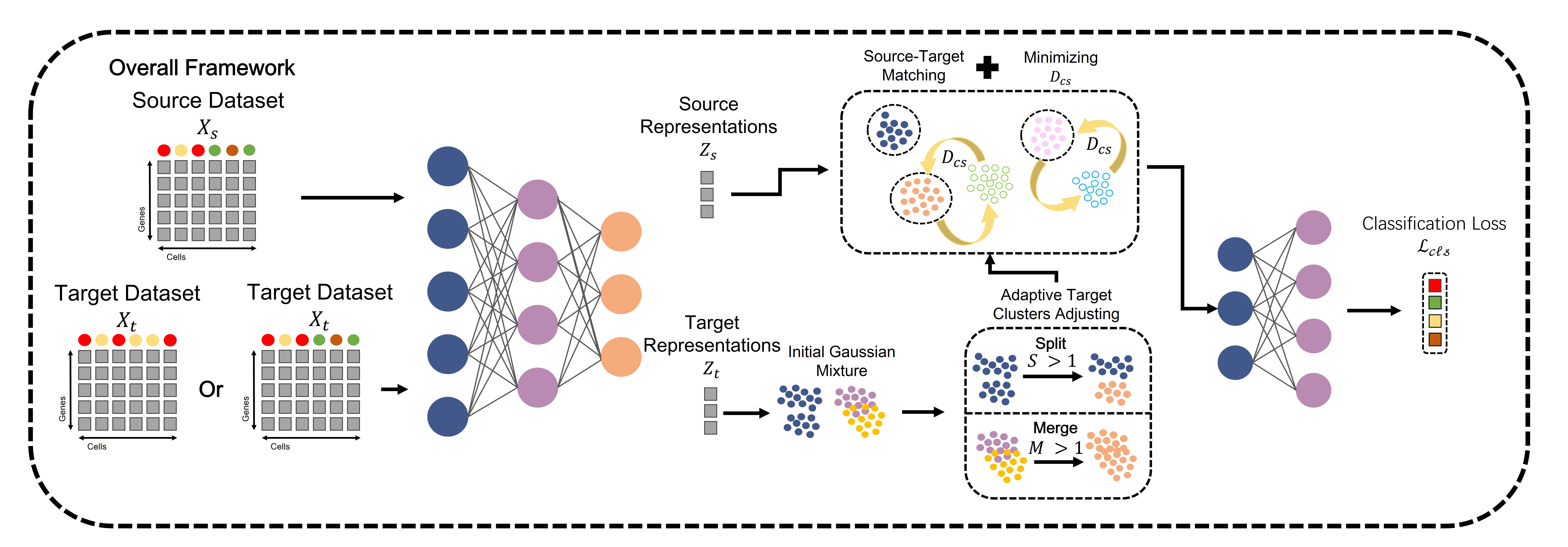}
\caption{Overall framework of ScNucAdapt}\label{Scnuc}
\end{figure}

The following passages are organized as follows. In Section \ref{methods}, the methods of ScNucAdapt are presented, and the descriptions of the datasets used and the evaluation metrics are included. In Section \ref{results}, the experimental results on classification accuracy between a set of scRNA-seq and snRNA-seq are presented, including an ablation experiment that demonstrates the effectiveness of each component in ScNucAdapt, as well as a sensitivity analysis and a runtime and memory scaling test. In Section \ref{Discussion}, the discussion is presented. Finally, in Section \ref{Conclusion}, the conclusion is presented.
\section{Materials and Method}\label{methods}

\subsection{Shared Source and Target encoder}
To extract features from both source and target datasets into a common label space, a shared encoder is used. The encoder is composed of two fully connected layers. The first layer transforms the input features into hidden units. The second layer reduces these features into a latent space, creating a compact representation that captures the most important patterns in the gene expression data.
\subsection{Dynamic clustering in Target data}
In this section, we will introduce the concept of dynamic clustering in the target dataset without the prior knowledge of the number of clusters, which was inspired by DeepDPM \cite {ronen2022deepdpm} for an unknown number of clusters for deep clustering, and PRAGA \cite {huang2025praga} for spatial multi-omics clustering.

Given the target datasets $X_{t} \in R^{n \times m}$, the target representations $Z_{t} \in R^{n \times m^{'}}$ are obtained in Eq. (\ref{extract}). Where $n$ represents the number of samples. $m$,$m^{'}$ represent the original number of features and the number of features after passing the shared encoder.
\begin{align}\label{extract}
    Z_{t} = MLP (X_{t})
\end{align}

After obtaining the representations of the target dataset, we set the initial cluster $C$ for the Gaussian mixture model to assign the target representations into $C$ clusters; note that $C$ doesn't represent the true number of clusters in the target representations or the true cell type labels in the source dataset, and would further be adjusted through a split and merge framework, which is performed through the Metropolis-Hastings framework \cite {hasting}.

The sample count and the target representations of each cluster are denoted by $N_{c}$ and $ Z^ {t} _ {c}$, respectively, where the subscript $c = 1,2,\dots, C$ denotes the cluster index. To enable dynamic adjustment of the total number of clusters, each cluster is further divided into two sub-clusters using Gaussian mixture. The corresponding sample count of the sub-clusters is represented as $N_{c,s}$, where $s \in \{1,2\}$ denotes the sub-cluster index. A splitting criterion is then defined for each cluster to determine whether it should be further partitioned, where the split is accepted with probability $\min{ (1, H_{s})}$. The hasting ratio $H_{s}$ is defined in Eq. (\ref{s}).

\begin{align}\label{s}
    H_{s} = \frac{\Gamma (N_{c,1})L (Z^{t}_{c,1};\nu,\kappa,m,\psi)\Gamma (N_{c,2})L (Z^{t}_{c,2};\nu,\kappa,m,\psi)}{\Gamma (N_{c})L (Z^{t}_{c};\nu,\kappa,m,\psi)}
\end{align}
where $\Gamma (\cdot)$ denotes the Gamma function, and $L (\cdot;\nu,\kappa,m,\psi)$ corresponds to the marginal likelihood evaluated under a Normal–Inverse–Wishart  (NIW) prior, parameterized by the hyperparameters $\nu$, $\kappa$, $m$, $\psi$. When $H_{s} > 1$, the original cluster is substituted with one of its derived subclusters, and the remaining subcluster is incorporated as an additional, distinct cluster shown in Eq. (\ref{yes}).
\begin{align}\label{yes}
    Z^{t}_{c} := Z^{t}_{c,1};Z^{t}_{C} := Z^{t}_{c,2} (C := C+1)
\end{align}
After splitting the clusters, we introduce a decision for merging criterion based on the clusters after splitting, where the merging is accepted with probability $\min{ (1, H_{m})}$, as shown in Eq.  (\ref{merge_decision}).
\begin{align}\label{merge_decision}
    H_{m (i,j)} = \frac{1}{H_{s (i,j)}} =  \frac{\Gamma (N_{i}+N_{j})L (Z^{t}_{i} \cup Z^{t}_{j};\nu,\kappa,m,\psi)}{\Gamma (N_{i})L (Z^{t}_{i};\nu,\kappa,m,\psi)\Gamma (N_{j})L (Z^{t}_{j};\nu,\kappa,m,\psi)}
\end{align}
Rather than exhaustively considering all possible merges in sequence, we limit the merge candidates for each cluster $N_{i}$ to its nearest neighbors $N_{j}$. If $H_{m (i,j)} > 1$, then the new merge cluster will replace the original two clusters, shown in Eq.  (\ref{merge}).
\begin{align}\label{merge}
    Z^{t}_{i} := \varnothing; Z^{t}_{j} := \varnothing ; Z^{t}_{C} := Z^{t}_{i} \cup Z^{t}_{j}  (C := C-1)
\end{align}

\subsection{Cauchy-Schwarz Divergence}\label{CS_divergence}
In this section, we introduce Cauchy-Schwarz Divergence \cite {yin2024domain} and the emperical estimator of $D_{CS} (p_{s} (z);p_{t} (z))$. Given the source representations of a certain class $Z_{s} \in R^{N \times m^{'}}$ and target representations of a certain cluster $Z_{t} \in R^{n \times m^{'}}$, where the source representations vectors $\{z_{i}^{s}\}_{i=1}^{N} \in Z_{s}$ and target representations vectors $\{z_{i}^{t}\}_{i=1}^{n} \in Z_{t}$, the Cauchy-Schwarz Divergence  (CS Divergence) is defined in Eq. (\ref{cs}).
\begin{align}\label{cs}
    D_{CS} (p_{s};p_{t}) = -\log{ (\frac{ (\int p_{s} (z)p_{t} (z)\mathrm{d}z)^{2}}{\int p_{s}^{2} (z)\mathrm{d}z\int p_{t}^{2} (z)\mathrm{d}z})}
\end{align}
Using kernel density estimation, we estimate the densities from finite samples defined in Eq. (\ref{cs1}) and Eq. (\ref{cs2}).
\begin{align}\label{cs1}
    \hat{p}_{s} (z) = \frac{1}{M}\sum^{M}_{i=1}\kappa_{\sigma} (z,z_{i}^{s})
\end{align}
\begin{align}\label{cs2}
    \hat{p}_{t} (z) = \frac{1}{N}\sum^{M}_{i=1}\kappa_{\sigma} (z,z_{i}^{t})
\end{align}
Here we chose gaussian kernel function as the kernel estimator $\kappa_{\sigma} (z,z^{'})=\exp{  (-\frac{\Vert z-z^{'}\Vert^{2}_{2}}{2\sigma^{2}})}$, where $\sigma$ is the bandwidth parameter that controls the smoothness of the kernel. The calculation results of $\int \hat{p}_{s}^2 (z)\mathrm{d}z$, $\int \hat{p}_{t}^2 (z)\mathrm{d}z$,\
$\int \hat{p}_{s} (z)\hat{p}_{t} (z)\mathrm{d}z$ are shown in Eq. (\ref{one}), Eq. (\ref{two}) and Eq. (\ref{three}).
\begin{align}\label{one}
    \int \hat{p}_{s}^{2} (z)\mathrm{d}z = \frac{1}{M^2}\sum^{M}_{i=1}\sum^{M}_{j=1}\kappa_{\sqrt{2}\sigma} (z_{j}^{s},z_{i}^{s})
\end{align}
\begin{align}\label{two}
    \int \hat{p}_{t}^{2} (z)\mathrm{d}z = \frac{1}{N^2}\sum^{N}_{i=1}\sum^{N}_{j=1}\kappa_{\sqrt{2}\sigma} (z_{j}^{t},x_{i}^{t})
\end{align}
\begin{align}\label{three}
    \int \hat{p}_{s} (x_{s})\hat{p}_{t} (x_{t})\mathrm{d}z = \frac{1}{MN}\sum^{M}_{i=1}\sum^{N}_{j=1}\kappa_{\sqrt{2}\sigma} (z_{j}^{t},z_{i}^{t})
\end{align}
By substituting Eq. (\ref{one})-Eq. (\ref{three}) into CS Divergence, we measure the divergence between source and target datasets as follows:
\begin{align*}
    D_{CS} (p_{s};p_{t}) &= \log (\frac{1}{M^2}\sum^{M}_{i,j=1}\kappa_{\sqrt{2}\sigma} (z_{j}^{s}-z_{i}^{s}))+\\
    &\log{ (\frac{1}{N^2}\sum^{N}_{i,j=1}\kappa_{\sqrt{2}\sigma} (z_{j}^{t}-z_{i}^{t}))}\\
    &-2\log (\frac{1}{MN}\sum^{M}_{i=1}\sum^{N}_{j=1}\kappa_{\sqrt{2}\sigma} (z_{j}^{t},z_{i}^{s}))
\end{align*}

\subsection{Source Class-Target Cluster Matching}\label{merg}
In this section, we will be introducing the Source Class-Target Cluster Matching after obtaining the CS divergence between source known cell classes and predicted target clusters. And further assign the matched pairs' CS Divergence to the final training loss.

Given the source representations vectors $Z_{s}$ and with $p$ labels, we could divide the source representations into $p$ subsets, noting as $Z_{s} = \{Z_{1,s},Z_{2,s},\dots,Z_{p,s}\}$. After performing dynamic clustering on the target representations on target representations vectors $Z_{t}$, we obtain $\hat{C}$ clusters $Z_{t} = \{Z_{1,t}, Z_{2,t},\dots, Z_{\hat{C},t}\}$. Then, in terms of CS divergence described in Section \ref{CS_divergence}, we calculate the CS divergence of each pair from the source classes and target clusters $Z_{i,s}$ and $Z_{j,t}$, which is $D_{CS} (p_{i,s}; p_{j,t})$, where $i = 1,2,\dots,p$ and $j = 1,2,\dots, \hat{C}$. To identify which subsets in the source representations best correspond to each cluster in the target representations, we selected those with the lowest CS divergence paired with each target cluster. Therefore, we could obtain the global merge decision, the total loss function for minimization on the mini-batched source and target dataset are shown in Eq. (\ref{cs_sum}), where $a_{i}$ represents the corresponding subsets that match the i-th target cluster.
\begin{align}\label{cs_sum}
    L_{cs} = \sum^{\hat{C}}_{i = 1}D_{CS} (p_{a_{i},s}^{b};p_{i,t}^{b})
\end{align}
\subsection{Shared Source and Target Classifier}
The shared source and target classifier operates on the latent representations produced by the encoder to predict the corresponding cell type. It consists of a single fully connected layer that maps directly to the output nodes, with the number of nodes equal to the number of cell types in the source dataset. This design allows the classifier to effectively translate the compact latent representations into accurate cell-type predictions while maintaining computational efficiency. The same classifier structure is consistently applied across all experiments without modification.
\subsection{The overview of ScNucAdapt}
In conclusion, ScNucAdapt consists of three parts. The framework is shown in Figure \ref{Scnuc}. The first is the shared encoder for the source and target datasets, aiming to extract representations in the same latent space. The second is the dynamic clustering in target representations. In this part, ScNucAdapt is responsible for clustering cell types in target datasets without giving prior knowledge on the number of clusters, and further adjusting through a split and merge framework. Then, we introduce CS Divergence and the rule for merging between the predicted target clusters and the source datasets. 

ScNucAdapt employs a two-stage training strategy, and the pseudocode is shown in supplementary S5 Fig. In the first stage, the encoder is trained for T warm-up epochs using only minibatch-based representations learning without clustering, thereby learning meaningful initial feature spaces. In the second stage, each epoch begins by applying GMM clustering and split/merge operation to the full learned representations of the target dataset, followed by source-target matching via argmin of Cauchy-Schwarz divergence between cluster distributions. The encoder is then updated via backpropagation on minibatches using the combined loss shown in Eq. (\ref{all_loss}) while treating the current cluster assignments as fixed. GMM clustering, split/merge operation, and source-target matching are recomputed every epoch to refine alignments as representations improve. To reduce computational time, we could also recompute every n epochs, but in our experiments, the operation is computed every epoch. Moreover, the total training loss on the selected source and target batch dataset consists of classification loss $L_{cls}$, which is the weighted cross-entropy loss, and the $L_{cs}$ described in section \ref{merg}. $\lambda$ represents the trade-off hyperparameter.
\begin{align}\label{all_loss}
    L = L_{cls} + \lambda*L_{cs}
\end{align}

\subsection{Datasets}\label{datasets}
Various scRNA-seq and snRNA-seq data are compiled from previous publications. Most datasets are preprocessed; datasets that require preprocessing are preprocessed using Scanpy  \cite {wolf2018scanpy}. We gathered cells from the bladder, kidney, the mouse cortex, and the frozen and fresh tumor tissues. 

For bladder cell types, the dataset is GSE267964 \cite {santo2025exploring}, which contains two subsets, Immune and Stromal. The datasets are preprocessed in advance and paired.

Moreover, we also collected unpaired scRNA-seq and snRNA-seq of kidney cell types
from different publications, GSE140989 \cite {menon2020single}, which is the scRNA-seq, and GSE121862 \cite {lake2019single}, which is the snRNA-seq. The cell type labels are gathered from a previous study on annotating cell types in kidneys from scRNA-seq and snRNA-seq using traditional machine learning methods. 

For frozen and fresh tumor cell types, the datasets were gathered from the GEO database under accession number GSE140819, which contains many types of frozen tumors of scRNA-seq and snRNA-seq. We collected the cell types from metastatic breast cancer (MBC) and Chronic lymphocytic leukemia (CLL), then we further preprocessed them by filtering cells and genes that have low counts. 

For mouse cortical cell types, the datasets are gathered from the GEO database under accession number GSE123454 \cite {bakken2018single}. All the cell type labels are collected from previous publications, with each cell annotated. 

The detailed statistics of the datasets are shown in Table \ref{stat_data}, including the number of samples, genes, and the number of unique cell types in each dataset.

\begin{table}[!t]
\caption{Statistical results of the datasets, including bladder cell types, kidney cell types, frozen and fresh tumor cell types, and mouse cortical cell types\label{tab1}}%
\begin{tabular*}{\columnwidth}{@{\extracolsep\fill}llll@{\extracolsep\fill}}
\toprule
Datasets & Cells  & Genes & Cell Types \\
\midrule
GSE267964-Immune (Sc)    & 1725 & 36387  &  9 \\
GSE267964-Immune (Sn)    &  369 & 36387 & 7  \\
GSE267964-Stromal (Sc)    & 7227 & 36387  & 8 \\
GSE267964-Stromal (Sn)    &5737 & 36387 & 8 \\
GSE140989 (Sc) & 20927&18743 & 13\\
GSE121862 (Sn) & 11684 & 18743 & 11\\
GSE123454 (Sc) & 463 &40023 &2\\
GSE123454 (Sn) & 463& 40023&2\\
GSE140819-CLL (Sc) & 2562& 33538& 3\\
GSE140819-CLL (Sn) & 2297& 33538& 2\\
GSE140819-MBC (Sc) & 5163 & 30316& 8\\
GSE140819-MBC (Sn) & 7260& 30316& 7\\

\bottomrule
\end{tabular*}
\label{stat_data}
\end{table}

Moreover, we provide the detailed adaptation settings between each dataset in Table \ref{setting}. These include a total of eight adaptation scenarios: four partial settings and four closed-set settings. Each configuration specifies the source–target dataset pairs, the shared and non-shared label spaces.
\begin{table}[!t]
\caption{Adaptation Settings Between datasets\label{setting}}%
\begin{tabular*}{\columnwidth}{@{\extracolsep\fill}llll@{\extracolsep\fill}}
\toprule
Datasets & Setting \\
\midrule
GSE267964-Immune (Sc)$\rightarrow$ GSE267964-Immune (Sn) & Partial\\
GSE267964-Stromal (Sc)$\rightarrow$GSE267964-Stromal (Sn) & Closed Set\\
GSE267964-Stromal (Sn)$\rightarrow$GSE267964-Stromal (Sc) & Closed Set\\
GSE140989 (Sc)$\rightarrow$GSE121862 (Sn) & Partial\\
GSE123454 (Sc)$\rightarrow$ GSE123454 (Sn) & Closed Set\\
GSE123454 (Sn)$\rightarrow$GSE123454 (Sc) & Closed Set\\
GSE140819-CLL (Sc)$\rightarrow$ GSE140819-CLL (Sn) & Partial\\
GSE140819-MBC (Sc)$\rightarrow$GSE140819-MBC (Sn) & Partial\\
\bottomrule
\end{tabular*}
\end{table}
\subsection{Evaluation Metrics}
To assess the performance of ScNucAdapt in cell type classification tasks, we evaluated its classification accuracy on datasets with known cell type annotations. The accuracy score quantifies the proportion of correctly predicted cell types among all predictions, providing a straightforward yet informative measure of model performance. Formally, accuracy is defined as shown in Eq. (\ref{acc}), where $y_{i}$ denotes the true label of the $i$-th sample, $\hat{y}_{i}$ represents the predicted class label for the same sample, and $n$ is the total number of samples in the dataset. The indicator function $1 (y_{i} = \hat{y}_{i})$ returns 1 if the predicted label matches the true label and 0 otherwise.

\begin{align}\label{acc}
    Acc (y_{i}, \hat{y}_{i}) = \frac{1}{n} \sum_{i=0}^{n-1} 1 (y_{i} = \hat{y}_{i})
\end{align}
Moreover, we also included the Macro-F1 score to evaluate model performance. Unlike overall accuracy, which can be biased toward majority classes, the Macro-F1 score provides a balanced assessment by treating all classes equally.

First, for each class $c \in \mathcal{C}$  (where $\mathcal{C}$ is the set of all classes), we compute class-specific precision and recall based on the true positives  ($TP_c$), false positives  ($FP_c$), and false negatives  ($FN_c$):

\begin{align}
\text{Precision}_c = \frac{TP_c}{TP_c + FP_c}
\end{align}

\begin{align}
\text{Recall}_c = \frac{TP_c}{TP_c + FN_c}
\end{align}

Precision measures the proportion of cells predicted to be class $c$ that are correctly assigned, reflecting the model's exactness. Recall measures the proportion of true class $c$ cells that were successfully retrieved, reflecting the model's completeness. The $F1$ score for each class is then defined as the harmonic mean of precision and recall, balancing the trade-off between the two:

\begin{align}
\text{F1}{c} = 2 \cdot \frac{\text{Precision}{c} \cdot \text{Recall}_{c}}{\text{Precision}_c + \text{Recall}_c}
\end{align}

Finally, the Macro-F1 score is calculated as the arithmetic mean of these per-class F1 scores across all classes:

\begin{align}
\text{Macro-F1} = \frac{1}{|\mathcal{C}|} \sum_{c \in \mathcal{C}} \text{F1}_c
\end{align}
\subsection{Experimental Setup}
The proposed method was evaluated against several existing classifiers that have been widely applied in single-cell transcriptomic analyses. Including SingleCellNet and ScMap. And also a popular domain adaptation method for the cross-batches cell type annotation method in scRNA-seq, ScAdapt \cite {zhou2021scadapt}. All the comparison methods are tuned to achieve the best performance and are performed on a research server equipped with an NVIDIA GeForce RTX 4090 GPU.

For initial clustering, we applied Gaussian Mixture Models with diagonal covariance matrices set to 'diag' in scikit-learn, and $C$ as the number of mixture components, which would be further adjusted.

In each experiment, the specific hyperparameters, including the width of the first hiddenlayer, the latent representations width, the early stopping, the trade-off hyperparameter, the learning rate, batchsize, and the epoch before performing initial Gaussian mixture clustering, and the initial cluster number K, are shown in Supplementary S1 Table. Moreover, the optimizer Adam and the hyperparameter $\sigma=5$ are fixed across datasets.

Given a set of counts $\{c_1, c_2, \ldots, c_n\}$ for $n$ classes, the weight $w_i$ for class $i$ is computed as:
\begin{align}
    w_i = \frac{1}{c_i}
\end{align}

Then, the weights in cross-entropy loss are calculated as:
\begin{align}
    \hat{w}_i = \frac{\frac{1}{c_i}}{\sum_{j=1}^{n} \frac{1}{c_j}} \times n
\end{align}
Where $c_i$ is the count for class $i$, $n$ is the total number of classes, and $\hat{w}_i$ is the final normalized weight for class $i$. Note that we've set a slightly higher weight for monocyte and plasma cell in bladder immune scRNA-seq to snRNA-seq. And on CLL, we set a slightly higher weight for the T cell.

The NIW hyperparameters are fixed across experiments and follow the settings from PRAGA and DeepDPM, where we used $ \kappa = 0.0001$, set $m$ to be the data mean, $\nu$ to be $K+2$, $K$ represents the initial clustering hyperparameter, and $\psi = I \times 0.005$, where $I$ denotes the identity matrix.
\section{Experimental Results}\label{results}
\subsection{Simulation Experiments}
We first generated simulated datasets using the R package Splatter \cite {splatter}, including both imbalanced and balanced datasets. Each dataset comprises two batches and five cell types. Following this, we performed two target-controlled experiments. In the first, we varied the batch effect strength by adjusting the batch.facScale parameter while keeping the number of cell types in the target dataset fixed at three. In the second experiment, we held the batch.facScale constant at one and varied the number of cell types present in the target data.
\begin{figure}[H]
\includegraphics[width=1\textwidth]{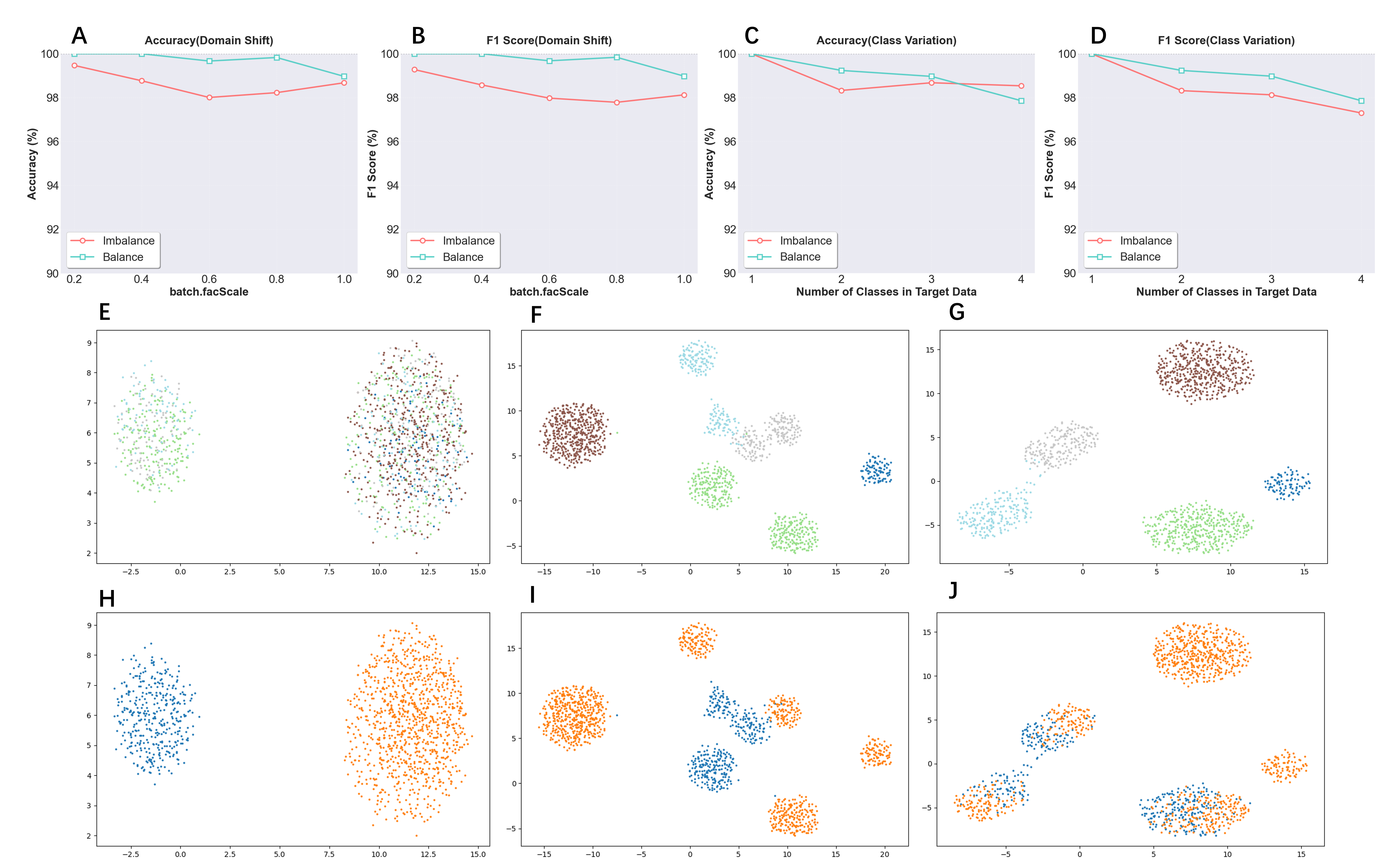}
\caption{Simulation Experiments using Splatter  (A) Accuracy on imbalanced and balanced simulated datasets domain shift experiments  (B) Macro f1-score on imbalanced and balanced simulated datasets domain shift experiments  (C) Accuracy on imbalanced and balanced simulated datasets class variation experiments  (D) Macro f1-score on imbalanced and balanced simulated datasets class variation experiments  (E) Uncorrected simulated dataset of cell types colored  (F) cell type representations before source classes and target clusters merging  (G) cell type representations after source classes and target clusters merging  (H) Uncorrected simulated dataset of batch colored  (I) batch representations before source classes and target clusters merging  (J) batch representations after source classes and target clusters merging.}\label{simulation}
\end{figure}
We provide UMAP visualizations shown in Figure \ref{simulation} to illustrate batch effects and cell type distributions under three conditions: before correction, at the onset of merging, and after complete merging.

The results show that across all conditions and evaluation metrics, ScNucAdapt consistently achieves robust performance. Furthermore, the UMAP visualizations shown in Figure \ref{simulation} on imbalanced simulated datasets with batch.facScale set to 1, which clearly demonstrates that ScNucAdapt effectively merges the correct cell types while exhibiting minimal negative transfer.
\subsection{ScNucAdapt Enables Accurate Cross-Annotation of Bladder and Kidney Cell Types Across scRNA-seq and snRNA-seq }\label{label1}
This section presents the classification performance of bladder and kidney cell types across domains between scRNA-seq and snRNA-seq data.

The experimental results shown in Table \ref{tab2} and in Supplementary S2 Table indicate that ScNucAdapt outperforms existing classifiers, which focus solely on scRNA-seq datasets. Moreover, ScAdapt. On the immune subset, which is a partial domain adaptation problem where scRNA-seq is the source data and snRNA-seq is the target data, ScNucAdapt achieves an accuracy of 91.05 and a macro-F1 Score 84.69, which performed better than ScAdapt's 90.24  (accuracy) and 82.37 (macro-f1), and outperformed SingleCellNet's 81.02 (accuracy) and 55.51 (macro-f1). On the Stromal datasets where scRNA-seq is the source data, and snRNA-seq is the target data, ScNucAdapt achieves an accuracy of 97.80 and 89.42 of macro-f1 under a closed set setting, performing better than ScAdapt's 96.95 (accuracy) and 80.00 (macro-f1), outperforming SingleCellNet's 91.92 (accuracy) and 63.82 (macro-f1). The same holds for the Stromal dataset, where snRNA-seq serves as the source data and scRNA-seq as the target data. ScNucAdapt achieves an accuracy score of 90.38 and 72.42 on macro-f1 score, outperforming ScAdapt's 89.98 (accuracy) and 71.61 (macro-f1), SingleCellNet's 86.61 (accuracy) and 66.96 (macro-f1), and ScMap's 87.02 (accuracy) and 63.72 (macro-f1). Interestingly, we found that domain adaptation methods often outperformed scRNA-seq classifiers. Visualization results using UMAP for the bladder tissue with three subsets are shown in Supplementary S3 Fig. While most scRNA-seq and snRNA-seq populations merged effectively, we observed a limitation in aligning stromal populations, where ScNucAdapt failed to distinguish between vein endothelial cells and general endothelial cells. This is attributable to the extremely limited training set for vein endothelial cells, which contained only four cells. This makes it a challenging scenario for any alignment method. Despite this, ScNucAdapt maintained the best overall performance among all compared methods in terms of both Accuracy and macro F1-score.

\begin{figure}[H]
\includegraphics[width=1\textwidth]{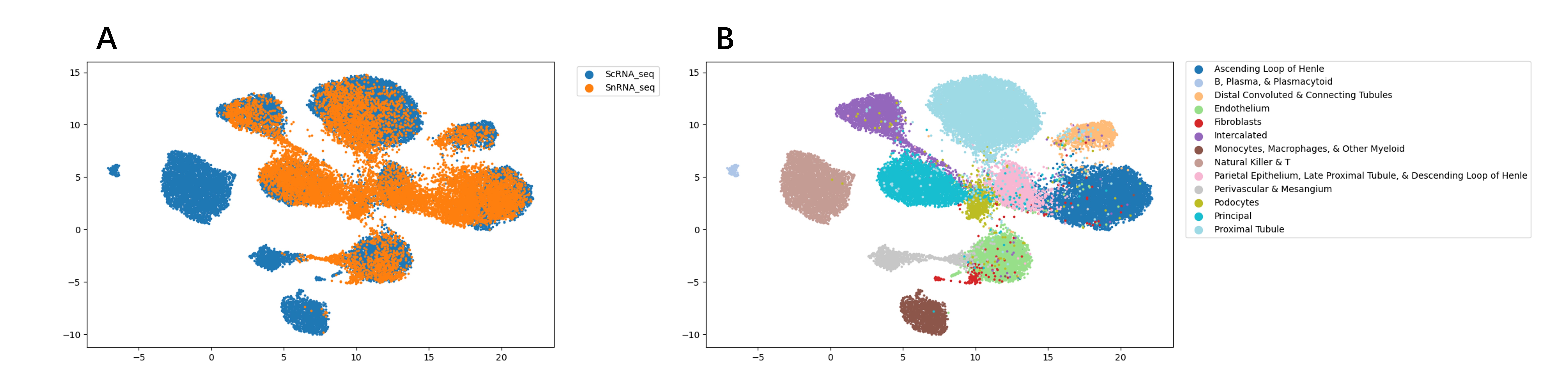}
\caption{Visualization result of scRNA-seq and snRNA-seq representations. Using UMAP on kidney tissue  (A) visualization result on scRNA-seq to snRNA batch representations  (B) visualization result on scRNA-seq to snRNA cell type representations}\label{kidney}
\end{figure}

On the unpaired datasets between scRNA-seq and snRNA-seq, which is under a partial setting, ScNucAdapt achieves an accuracy of 87.23 and a macro-f1 of 81.5, outperforming other methods, including ScAdapt. The visualization results of UMAP are shown in Figure \ref{kidney}, where most scRNA-seq and snRNA-seq populations are well-merged across batches while maintaining clear separation by cell type. These results indicate that not only can ScNucAdapt handle distributional differences between scRNA-seq and snRNA-seq, but also under partial settings of unpaired scRNA-seq and snRNA-seq.

All the experimental results show that ScNucAdapt is a robust method in cross-domain annotation between scRNA-seq and snRNA-seq in bladder and kidney cell types. 
\begin{table*}[t]
\caption{Classification accuracy of target datasets on bladder and kidney cell types.\label{tab2}}
\tabcolsep=0pt
\begin{tabular*}{\textwidth}{@{\extracolsep{\fill}}lcccc@{\extracolsep{\fill}}}
\toprule%
Datasets & \multicolumn{1}{@{}c@{}}{ScMap} & \multicolumn{1}{@{}c@{}}{SingleCellNet}& \multicolumn{1}{@{}c@{}}{ScAdapt} & \multicolumn{1}{@{}c@{}}{ScNucAdapt} \\

\midrule
GSE267964-Immune (Sc)$\rightarrow$GSE267964-Immune (Sn) & 75.06& 81.02 & 90.24 & \textbf{91.05} \\
GSE267964-Stromal (Sc)$\rightarrow$GSE267964-Stromal (Sn) & 79.58 & 91.92 & 96.95 & \textbf{97.80}\\
GSE267964-Stromal (Sn)$\rightarrow$GSE267964-Stromal (Sc) & 87.02& 86.61& 89.98&\textbf{90.38} \\
GSE140989 (Sc)$\rightarrow$GSE121862 (Sn) & 86.04&70.58 &84.01 &\textbf{87.23}\\
\bottomrule
\end{tabular*}

\end{table*}
\begin{table*}[t]
\caption{Classification accuracy of target datasets on frozen and fresh tumor cell types.\label{tab3}}
\tabcolsep=0pt
\begin{tabular*}{\textwidth}{@{\extracolsep{\fill}}lcccccc@{\extracolsep{\fill}}}
\toprule%
Datasets & \multicolumn{1}{@{}c@{}}{ScMap} & \multicolumn{1}{@{}c@{}}{SingleCellNet}& \multicolumn{1}{@{}c@{}}{ScAdapt} & \multicolumn{1}{@{}c@{}}{ScNucAdapt}  \\

\midrule
GSE140819-CLL (Sc)$\rightarrow$  GSE140819-CLL (Sn) & 97.64 & 93.07 & 96.99 & \textbf{98.39}\\
GSE140819-MBC (Sc)$\rightarrow$GSE140819-MBC (Sn) & 84.88 &64.82 & 94.17& \textbf{95.39}\\
\bottomrule
\end{tabular*}

\end{table*}
\begin{table*}[t]
\caption{Classification accuracy of target datasets on mouse cortical cell types.\label{tab4}}
\tabcolsep=0pt
\begin{tabular*}{\textwidth}{@{\extracolsep{\fill}}lcccc@{\extracolsep{\fill}}}
\toprule%
Datasets & \multicolumn{1}{@{}c@{}}{ScMap} & \multicolumn{1}{@{}c@{}}{SingleCellNet}& \multicolumn{1}{@{}c@{}}{ScAdapt} & \multicolumn{1}{@{}c@{}}{ScNucAdapt} \\

\midrule
GSE123454 (Sc)$\rightarrow$ GSE123454 (Sn) & 98.48 & 99.56& 99.56 & \textbf{99.78}\\
GSE123454 (Sn)$\rightarrow$GSE123454 (Sc) & 99.13 & \textbf{100.00} & \textbf{100.00} &\textbf{100.00}\\
\bottomrule
\end{tabular*}

\end{table*}

\subsection{ScNucAdapt Supports Reliable Cross-Annotation of Fresh and Frozen Tumor Cells Between scRNA-seq and snRNA-seq }
This section presents the classification performance of cell types in fresh and frozen tumors. As described in section \ref{datasets}, we chose two types of tumors, metastatic breast cancer and chronic lymphocytic leukemia.

The experimental results presented in Table \ref{tab3} and Supplementary S2 Table show that, under a partial adaptation setting for metastatic breast cancer, where scRNA-seq serves as the source data and snRNA-seq as the target data, ScNucAdapt performed better than ScAdapt's 94.17 (accuracy) and 74.89 (macro-f1), achieving an accuracy of 95.39 and a macro-f1 score of 76.16. Both methods outperform other comparison methods. 

Moreover, the cross-domain annotation from scRNA-seq to snRNA-seq on chronic lymphocytic leukemia, ScNucAdapt, achieved an accuracy of 98.39 and a macro-f1 of 94.47, which performed better than existing scRNA-seq cell type classifiers and ScAdapt. These results demonstrate that ScNucAdapt is effective under partial adaptation settings and can reliably handle cross-domain annotation tasks.

\begin{figure}[H]
\includegraphics[width=1\textwidth]{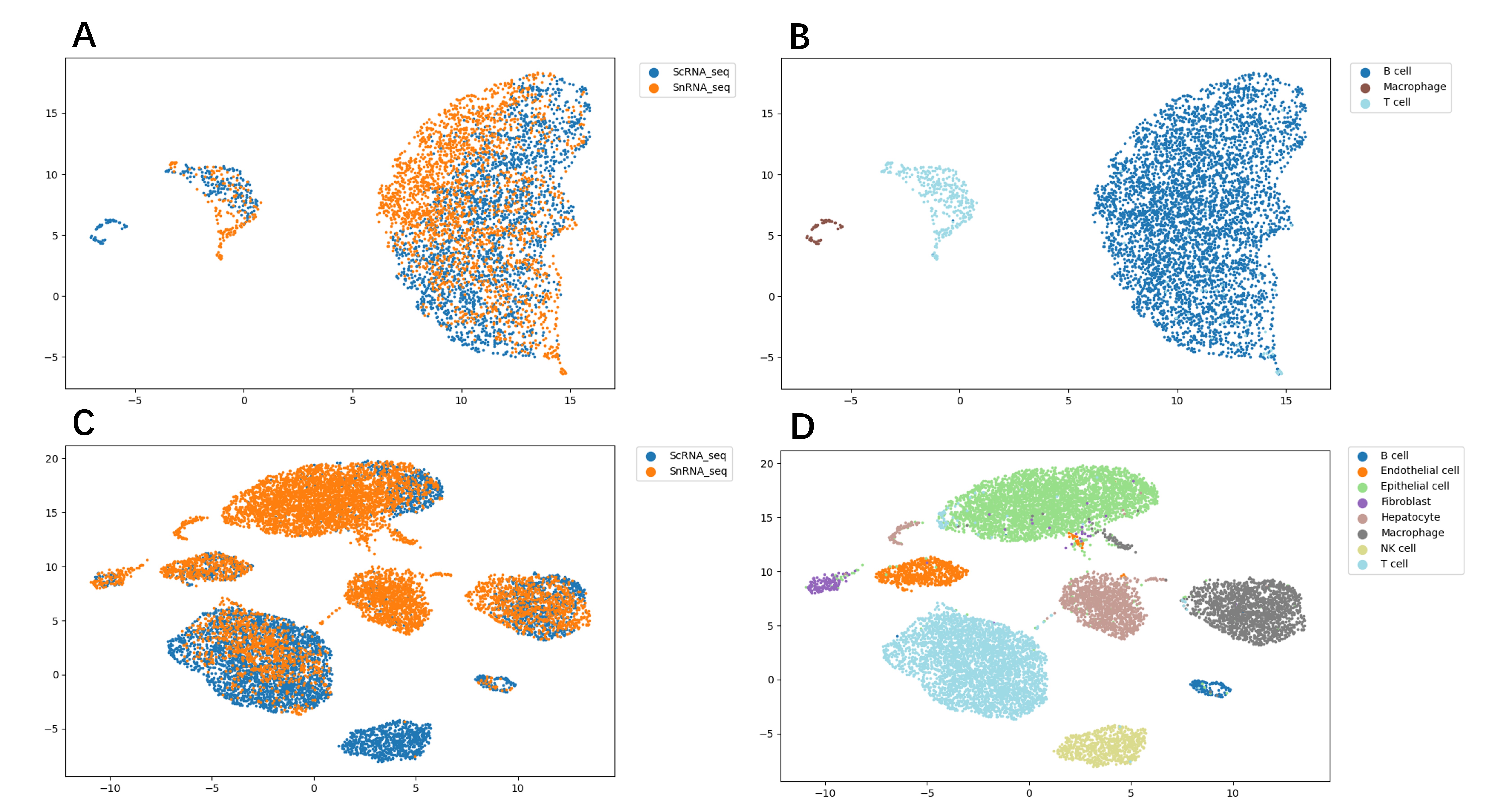}
\caption{Visualization result of scRNA-seq and snRNA-seq representations. Using UMAP on frozen and fresh tumor tissue  (A) visualization result on scRNA-seq to snRNA batch representations using CLL (B) visualization result on scRNA-seq to snRNA cell type representations using CLL  (C) visualization result on scRNA-seq to snRNA batch representations using MBC (D) visualization result on scRNA-seq to snRNA cell type representations using MBC.}\label{frozen}
\end{figure}

Figure \ref{frozen} presents the UMAP visualization results, illustrating that cells from scRNA-seq and snRNA-seq are well-mixed after integration, yet remain distinctly separated according to their cell type identities.
\subsection{ScNucAdapt Enables Cross-Annotation of Mouse Cortical Cell Types Across scRNA-seq and snRNA-seq}

We have also included the cross-domain cell type annotation on mouse cortical cell types. The experimental settings are identical to those described in the previous sections.

The experimental results shown in Table \ref{tab4} and supplementary S3 Table revealed that the cross-annotation from scRNA-seq to snRNA-seq, ScNucAdapt, achieves an accuracy score of 99.78, which performed better than SingleCellNet, ScMap, and scAdapt. On the cross-annotation from snRNA-seq to scRNA-seq, ScNucAdapt, ScAdapt, and SingleCellNet achieved an accuracy of 100.00, while ScMap achieved 99.13 (accuracy) and 97.99 (macro-f1).

As shown in Supplementary S4 Fig, the UMAP embeddings demonstrate successful integration, with cells from both scRNA-seq and snRNA-seq mixing effectively while maintaining clear separation by cell type.
\subsection{Ablation Experiments}

In this section, we conducted an ablation study to evaluate the contribution of each major component of the proposed ScNucAdapt framework. Two core modules were examined: 

\begin{itemize}
    \item The use of CS divergence to measure and minimize distributional discrepancies between the source subset and the target clusters, thereby aligning their feature distributions.
    \item The dynamic cluster selection mechanism, which identifies clusters within the target domain without requiring prior knowledge of their number.
\end{itemize}

We hypothesize that removing the CS divergence would weaken the model’s ability to generalize across domains, while omitting the dynamic cluster selection would impair the model’s capacity to adapt effectively to target data due to insufficient structural guidance.

\begin{figure}[H]
\includegraphics[width=1\textwidth]{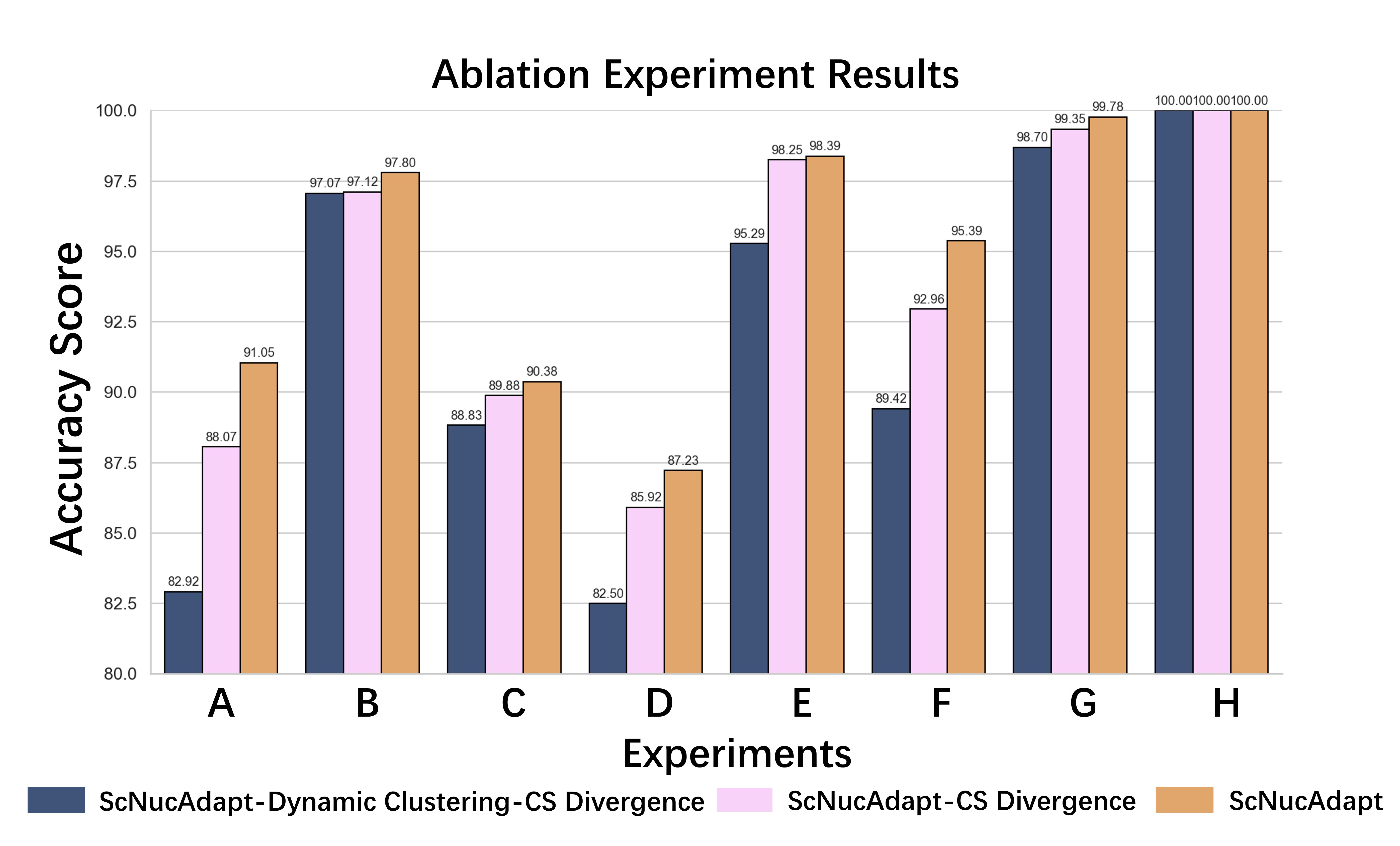}
\caption{Ablation experiments conducted on 8 cross domain classification.  (A) Ablation analysis on GSE267964-Immune (Sc)$\rightarrow$GSE267964-Immune (Sn)  (B) Ablation analysis on GSE267964-Stromal (Sc)$\rightarrow$GSE267964-Stromal (Sn)  (C) GSE267964-Stromal (Sn)$\rightarrow$GSE267964-Stromal (Sc)  (D) Ablation analysis on GSE140989 (Sc)$\rightarrow$GSE121862 (Sn)  (E) Ablation analysis on GSE140819-CLL (Sc)$\rightarrow$  GSE140819-CLL (Sn)  (F) Ablation analysis on Sensitivity analysis on GSE140819-MBC (Sc)$\rightarrow$  GSE140819-MBC (Sn)  (G) Ablation analysis on GSE123454 (Sc)$\rightarrow$ GSE123454 (Sn)  (H) Ablation analysis on GSE123454 (Sn)$\rightarrow$ GSE123454 (Sc)}\label{ablation}
\end{figure}

To test these hypotheses, ablation experiments were performed on five cross-domain cell-type annotation tasks, including datasets on bladder cell types, kidney cell types, and tumor cell types. The results shown in Figure \ref{ablation} reveal that excluding either component substantially decreases generalization performance, indicating that both CS-based distributional alignment and dynamic cluster selection are critical for robust cross-domain annotation. 

In particular, removing the dynamic cluster selection module, while retaining the CS divergence, still led to a noticeable decline in accuracy, underscoring the complementary role of adaptive clustering in enhancing ScNucAdapt’s ability to generalize between scRNA-seq and snRNA-seq domains. These findings highlight the importance of both proposed components in mitigating modality-specific distributional differences and achieving stable cross-domain cell-type classification.

\subsection{Sensitive Analysis}

In the previous section, we introduced the hyperparameter $C$, which is the initial cluster for the Gaussian mixture model. However, the prior knowledge of the real number of clusters in the target dataset is unknown. Our proposed method is capable of dynamically selecting the appropriate number of clusters after being given the hyperparameters. Therefore, we conducted a sensitivity analysis on the hyperparameter $C$ to see whether there is extreme fluctuation in the performance when different hyperparameters $C$ are given.

\begin{figure}[H]
    \begin{center}
        \includegraphics[width=1\textwidth]{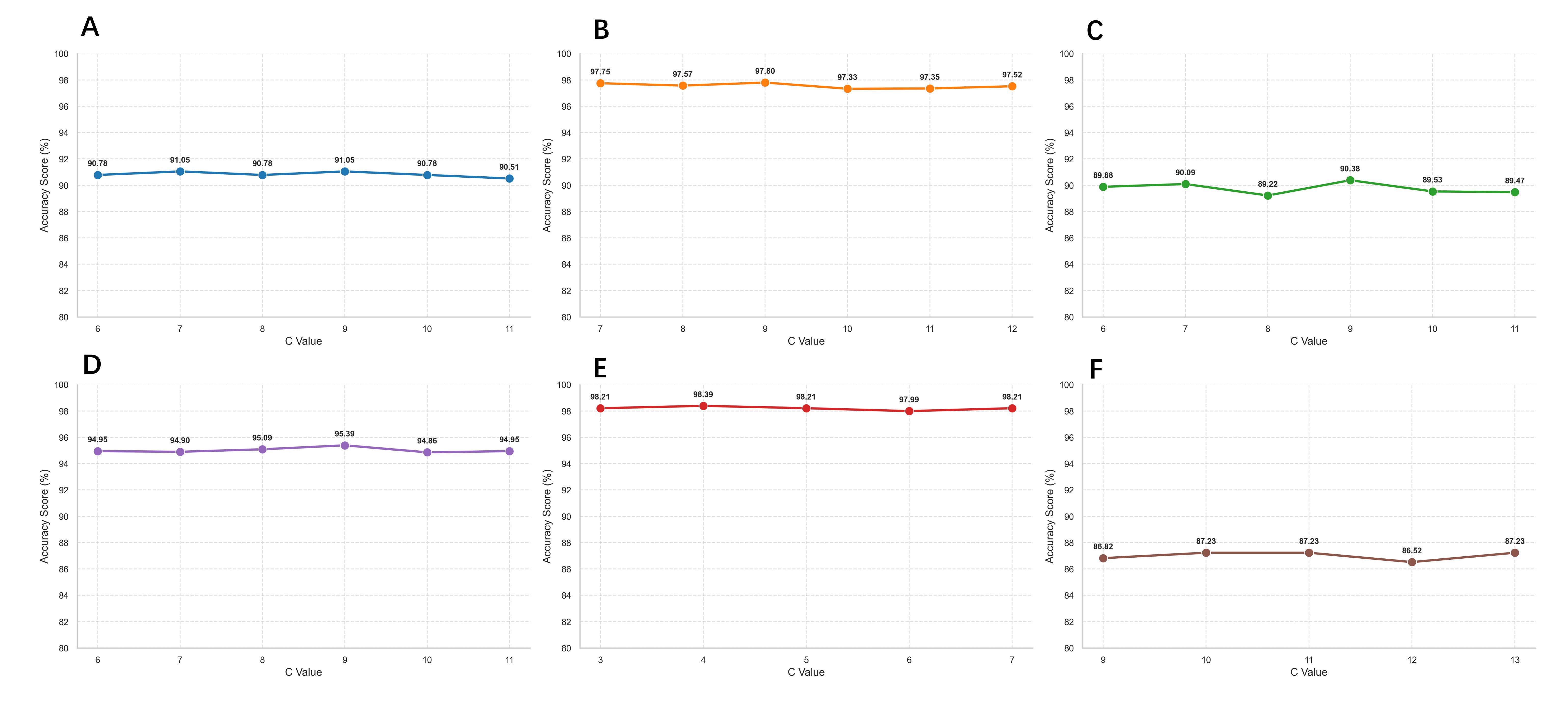}
\caption{Hyperparameter Sensitive analysis on $C$.  (A) Sensitivity analysis on GSE267964-Immune (Sc)$\rightarrow$GSE267964-Immune (Sn)  (B) Sensitivity analysis on GSE267964-Stromal (Sc)$\rightarrow$GSE267964-Stromal (Sn)  (C) Sensitivity analysis on GSE267964-Stromal (Sn)$\rightarrow$GSE267964-Stromal (Sc)  (D) Sensitivity analysis on GSE140819-MBC (Sc)$\rightarrow$  GSE140819-MBC (Sn)  (E) Sensitivity analysis on GSE140819-CLL (Sc)$\rightarrow$  GSE140819-CLL (Sn)  (F)  Sensitive analysis GSE140989 (Sc)$\rightarrow$GSE121862 (Sn)}\label{sensitive}
    \end{center}
\end{figure}

A total of six cross-domain cell type annotation experiments are included in the analysis. The experimental results shown in Figure \ref{sensitive} show that on most occasions, ScNucAdapt is insensitive to the hyperparameters. However, we noticed that there's a small fluctuation when conducting sensitive analysis on the immune bladder cell type from scRNA-seq to snRNA-seq, but the performance wasn't significantly degraded. The results suggest that ScNucAdapt can automatically adapt to diverse datasets without the need for extensive hyperparameter tuning, thereby improving its applicability in real-world cross-domain annotation between scRNA-seq and snRNA-seq.

Moreover, we conducted sensitivity analysis on the trade-off hyperparameter $\lambda$ following the six cross-domain cell type annotation experiments. The experimental results shown in Supplementary S1 Fig indicate that ScNucAdapt is not sensitive to the changes of the trade-off hyperparameter across a wide range of values. The model maintained stable performance for $\lambda$, demonstrating that the proposed method achieves consistent domain adaptation without requiring extensive hyperparameter tuning.
\subsection{Runtime and Peak GPU Memory Scalability Tests}
To assess the runtime and peak GPU memory of ScNucAdapt, we performed scaling experiments on simulated single-cell RNA-seq datasets generated with Splatter. We simulated four datasets with approximately 2,000, 5,000, 10,000, and 20,000 cells  (10,000 genes, batch.facScale 0.8), which, after removing two cell types from the target batch, resulted in final sizes of 1,613, 4,013, 7,985, and 16,011 cells, respectively. The results shown in Supplementary S2 Fig show that interestingly, memory consumption scaled linearly from 0.065GB to 0.413GB across these dataset sizes, with peak memory remaining stable during encoder updates due to minibatch-based backpropagation. Runtime per epoch increased from 45.7 seconds to 1,436.4 seconds, with the computational bottleneck being the GMM clustering and split-merge operations performed on the full dataset each epoch. Therefore, we could rerun the clustering and matching every n epochs to reduce runtime.
\section{Discussion}\label{Discussion}
In this paper, our study fills the research gap on the cross-domain annotation between scRNA-seq and snRNA-seq, and also addresses the distributional and cell composition differences between the two types of datasets. The experimental results show the robustness of ScNucAdapt on the cross-domain annotation between scRNA-seq and snRNA-seq, and further ablation experiments have proved the effectiveness of the proposed components in ScNucAdapt. Moreover, insensitive to the hyperparameters that need manual controls on the initial clusters.

While ScNucAdapt is proposed for the cross-domain annotation between scRNA-seq and snRNA-seq and shows robustness, several problems and questions could be addressed in future work. 

First, label noises  \cite {chen2025timestamp} that existed in the source datasets could degrade the performance by introducing unreliable supervision signals. These mislabeled samples may hinder domain alignment and reduce the accuracy of downstream annotation.

Second, an exciting direction for future research lies in novel cell type discovery across the target dataset \cite {shi2024scadca}. Current cross-domain annotation frameworks rely heavily on existing cell-type labels and may overlook previously uncharacterized or rare cell populations that are domain-specific. Therefore, future work must integrate Open-Set Domain Adaptation \cite {panareda2017open} or universal domain adaptation frameworks \cite {you2019universal}, as these are specifically designed to handle both shared and private label sets, allowing models to flag target-domain-specific cells as "unknown" for further validation.

A more challenging but realistic direction involves scenarios where the gene sets differ substantially between scRNA-seq and snRNA-seq. This moves the problem into the realm of Heterogeneous Domain Adaptation \cite {liu2020heterogeneous}, where the feature spaces themselves are mismatched. Future frameworks capable of projecting heterogeneous gene sets into a common latent space would vastly expand the flexibility and applicability of cross-domain annotation.

Moreover, due to high sparsity and high-dimensional spaces occurring in both scRNA-seq and snRNA-seq, ScNucAdapt tends to overfit, although ScNucAdapt tends to generalize well compared to existing methods on cross-domain annotation between scRNA-seq, there's still room for improvements on the performance. Therefore, future direction could focus on developing algorithms to prevent overfitting and achieve better generalization on target datasets \cite {he2023addressing}. 

Finally, imbalanced cell type distributions within each domain could hinder generalization. Overrepresented cell types may dominate the training process, causing the model to underperform on rare populations. Future methods could focus on addressing these within-domain imbalances to improve robustness and cross-domain performance  \cite {wang2023titok}.
\section{Conclusion}\label{Conclusion}
In this study, we introduced ScNucAdapt, a novel cross-domain annotation framework designed specifically for transferring cell type labels between paired or unpaired scRNA-seq and snRNA-seq datasets. To the best of our knowledge, this is the first method to address the unique challenges of cross-annotation across these two sequencing protocols. ScNucAdapt tackles both distributional differences and label space mismatches between source and target domains through three key components: a shared encoder that projects both datasets into a common latent space, a dynamic clustering mechanism that adaptively identifies the unknown number of cell types in the target data via split-merge operations, and a Cauchy-Schwarz divergence-based matching strategy that aligns source classes with target clusters while minimizing negative transfer from non-shared cell types.

Extensive experiments on eight cross-domain annotation tasks spanning bladder, kidney, tumor, and mouse cortical tissues demonstrated that ScNucAdapt consistently outperforms existing methods, including scRNA-seq classifiers and domain adaptation baselines, in both accuracy and macro F1-score. The framework proves effective under both closed-set and partial-set scenarios, maintaining robust performance even when target label spaces are subsets of the source. Ablation studies confirmed the necessity of each proposed component, while sensitivity analyses showed that ScNucAdapt is robust to hyperparameter choices, requiring minimal tuning in practice.

Our scalability analysis on simulated datasets confirmed that ScNucAdapt exhibits linear memory scaling and manageable runtime for datasets up to 16,000 cells, with the primary computational bottleneck being the GMM clustering and split-merge operations performed each epoch. For larger datasets, we recommend reducing the frequency of these operations or exploring approximate clustering variants.

Despite these strengths, several promising directions remain for future work. These include handling label noise in source annotations, extending the framework to discover novel cell types in target domains through open-set or universal domain adaptation, addressing heterogeneous feature spaces where gene sets differ substantially between datasets, mitigating overfitting in high-dimensional sparse spaces, and developing strategies to better handle imbalanced cell type distributions. Addressing these challenges will further enhance the applicability and robustness of cross-domain annotation methods in real-world single-cell and single-nucleus studies.

In summary, ScNucAdapt provides a powerful and flexible solution for integrating and annotating scRNA-seq and snRNA-seq data, enabling more consistent and reliable interpretation of cellular identities across experimental protocols and tissue conditions.
\section*{Competing interests}
The authors declare that there are no competing interests.
\section*{Acknowledgments}
This work is supported in part by funds from the National Natural Science Foundation of China (62131004, 62531002, 62306051, 62481540175, 62276035), the Taishan Scholars Foundation of Shandong Province (tsqn202507225), and the Natural Science Foundation of Chongqing  (CSTB2025NSCQ-GPX0857). The Fundamental Research and the Scientific and the Technological Research Program of Chongqing Municipal Education Commission  (KJQN202300718).

\section*{Code availability}
The source code is located at https://github.com/OPUS-Lightphenexx/ScNucAdapt.

\section*{Data Availability Statement}
The data used in this study are available in Gene Expression
Omnibus  (GEO) with accession numbers GSE267964, GSE140989, GSE121862, GSE123454, GSE140819.

\bibliography{reference}


\end{document}